# Distributed approach for the indoor deployment of wireless connected objects by the hybridization of the Voronoi diagram and the Genetic Algorithm

**Wajih Abdallah**
UT2J, CNRS-IRIT (RMESS), University of Toulouse, Toulouse, France
wajih.abdallah@irit.fr

ISAM Gafsa, Dept. of Design, University of Gafsa ,Tunisia
wajih.abdallah@uvt.tn

**Sami Mnasri**
CNRS-IRIT (RMESS), University of Toulouse, Toulouse, France
Sami.Mnasri@irit.fr
University of Tabuk, Community College, Dept. of computer sciences, Tabuk, KSA
smnasri@ut.edu.sa

**Thierry Val**
UT2J, CNRS-IRIT (RMESS), University of Toulouse, Toulouse, France
val@irit.fr

**Abstract:**

IoT data collection networks have recently become one of the important research areas due to their fundamental role and wide application in many domains. The establishment of networks of objects is based essentially on the deployment of connected objects to process the collected data and transmit them to the various locations. Subsequently, a large number of nodes must be adequately deployed to achieve complete coverage. This manuscript introduces a distributed approach, which combines the Voronoi Diagram and the Genetic algorithm(VD-GA), to maximize the coverage of a region of interest. The Voronoi diagram is used to divide region into cells and generate initial solutions that present the positions of the deployed IoT objects. Then, a genetic algorithm is executed in parallel in several nodes to improve these positions.

The developed VD-GA approach was evaluated on an experimental environment by prototyping on a real testbed utilizing M5StickC nodes equipped with ESP32 processor. The experiments show that the distributed approach provided better degree of coverage, RSSI, lifetime and number of neighboring objects than those given by the original algorithms in terms of the suggested distributed Genetic-Voronoi algorithm outperforms the centralized one in terms of speed of computing.

.

## 1 Introduction

In recent years, IoThas been applied in several areas such as cities, warehouses, buildings, hospitals, universities and companies [1, **2**].An IoT data collection network is generally composed of a large number of multi-functional connected objects called nodes which form the wireless sensor network. The latter can gather and send information. It also allows controlling or detecting objects directly via the used network technology used, establishing a closer connection between the real world and computer networks and improving network performance metrics[**3**, 4].

The basic elements of a data collection network are sensor nodes made up of detection unit, processing unit, storage unit and communication unit [5]. The network operates according to three phases: detection, processing and transmission of datato the receiver through a number of intermediate nodes[6]. The detection unit of a sensor node interfaces with its physical environment to accumulate the required information. In fact, the transmission of data and control packets through the network is performed by the communication unit of a node [7]. Sensor nodes are also equipped with a processing unit that can process data. Generally, this unit consists of a microprocessor or microcontroller. It is also linked to a memory unit used to store a set of instructions and detected data [8].

The main objectives of using IoT data collection networks are to monitor and control the target within the region of interest [9,10], on the one hand, and determine any physical action (e.g. light, heat, sound, etc.), on the other hand The sensor node detects, processes, and convert these observations into a compatible format. IoT collection networks are implemented for many uses and applications in various fields such as military applications [**11**], industrial applications [12], commercial applications [13], home applications [14], healthcare applications [15] and environmental applications [16].

Many issues, such as limited storage in a node, routing problem and how to develop a good routing algorithm that guarantees efficient data transfer between sensor nodes, are encountered in IoT data collection networks. In fact, the performance of the latter is affected by some parameters like the localisation of a sensor node in a region of interest and the deployment of nodes. The authors [17] pointed out that the latter is an essential design aspect in setting up a network because it reflects the detection and control ability as well as the cost of an IoT network. It also affects almost all its performance metrics like coverage, network lifetime, and connectivity between nodes [18].

The methods of deployment in a network of connected objects can be classified into two categories: deterministic and random. In fact, random deployment is carried distributing sensor nodes randomly over the whole region of interest (RoI) [19, 20]. This type of deployment is usually used in applications where the area to be monitored is too large or where the region of interest is inaccessible due to adverse and hazardous environmental conditions In this case, objects can be dispersed from an aircraft or by using multi-robot systems [21]. To attain complete coverage, the random deployment technique requires more sensor nodes than the intended deployment [20].

An optimal deployment of sensor nodes has two main objectives. The first aim is to maximize the coverage of a given region of interest, while the second objective consists in minimizing the number of sensor nodes to be deployed [22,**23**]. This type of problem is classified as an NP-hard optimization problem [24].

Many stochastic optimization algorithms have been designed to solve different optimization problems. Among these algorithms commonly used in the deployment of data collection networks, we can mention: the Genetic Algorithm (GA) [25,26], the Particle Swarms Optimization (PSO) [27,**28**], the Artificial Bee Colony (ABC) [ 29], Ant Colony Optimization (ACO) [30], Fruit Fly Optimization (FOA) [31], Bacterial Food Search Optimization algorithm (BFO) [32], Group Search Optimization (GSO) [33], Harmony Search Algorithm (HSA) [34], Charged Search System (CSS) [35], Rivers Formation Dynamics (RFA) [36]. In previous academic studies, these algorithms were combined and hybridized with other paradigms to enhance the performance of deployment in terms of convergence and precision rate.

In this paper, we propose a distributed approach based on the hybridization of a geometric deployment method, used in the Voronoi Diagram, and a genetic algorithm. This processing is performed in several nodes in parallel to reach the desired solution in a shorter time compared to the centralized approach. This hybridization aims at positioning the nodes in a network and, subsequently, to maximize the coverage of a RoI with the minimum number of nodes. The Voronoi Diagram (VD) presents the starting point of our work that generates an initial population with a random deployment of objects. This task is performed on a single M5StickC node. Then, the initial population will be distributed over all the other M5StickC nodes forming the data collection network. Thus, a genetic algorithm is executed in each of these nodes to optimize deployment.

The work presented in this article has three main contributions:

1. A hybridization between distributed VD and GA is introduced, with the aim of improving the coverage rate in data collection networks.
2. The introduced distributed VD-GA approach was evaluated in an experimental environment by prototyping on real testbeds using M5StickC nodes (This type of nodes is used for the first time for addressing the deployment issues).
3. Evaluate and compare the found result with a centralized VD-GA approach and other deployment paradigms.

This paper is divided into four sections: Section 2 overviews some research works related to deployment optimization. Section 3 presents our distributed approach based on the hybridization of the VD algorithm and the genetic algorithm (GA) proposed for the deployment of objects in

IoT collection networks. Section 4 illustrates and discusses the obtained results and compares the performance of the suggested approach with that of our centralized approach [37] is performed on a single node.

## 2 Relatedworks

### 2.1 Hybridization of DV and optimization algorithms

The heuristic method that is widely used for the deployment and optimization in IoT data collection networks is optimization by genetic algorithms (GA) applied to improve the positions of nodes in order to maximize the region of interest coverage and, subsequently, extend the life of the network.This technique is also utilized to reduce energy consumption. Other commonly-used heuristics include optimization by particle swarms (PSO), optimization by ant colonies (ACO), which is generally employed to optimize routing paths between nodes. Voronoi-based approaches used in many studies to maximize the coverage in an area of interest by detecting coverage holes in data collection network.

In this section, we present some studies that proposed hybridization of VD and optimization algorithms in different applications and contexts. Indeed, several works combined Voronoi-based approaches with other nature-inspired algorithms such as ant colonies optimization, particles warms optimization and genetic algorithms to maximize the coverage of a region of interest, as shown in Tab.1.

Table1. Hybridization of Voronoi and other optimization algorithms (ACO), (PSO) and (GA)

|   | Hybridizationmethods | Functioning | Impact |
|---|---|---|---|
| Voronoi + ACO [30, 38,39] | Voronoi: determine all the possible paths in a network | Assign weight values to the Voronoi edges for the routing path search. | -Nodes distribution -Weight values - Evaluation fonction. |
|  | ACO: identify the shortest path among all the paths generated by Voronoi |  |  |
| Voronoi + PSO [27, 40, 41] | Voronoi: detect coverage holes in a region of interest | Generate virtual points from the detected holes. The detection range issometimes changed. | -Virtual points -Nodes localisation -Nodes speed -Best local or global solution. |
|  | PSO: generate Voronoi vertices (virtual points) or random endpoints to reduce power consumption and maximize the network lifetime. |  |  |

| Voronoi + GA [42, 43,44] | Voronoi: detect coverage holes in a region of interest | Detection of holes, then repositioning of the nodes to be developed with the modification of the level of nodes distribution | -Impact on the evaluation function. -Impact on the hole coverage rate -Adding sometimes other mobile nodes to the detected holes |
|---|---|---|---|
| | GA: generate new candidate solutions (i.e. new node locations) tomaximize the coverage of a region of interest. | | |

Many works are based on the hybridization of Voronoi and ant colonies optimization (ACO) for node deployment in a network. Moreover, VD was primarily used to optimize path planning to ensure efficient communication between the different nodes of the network. The most important roles of ACO are to: i) find connection paths in a network and ii) to search and adapt tothe shortest path [45].

Hybridizations of Voronoi and particle swarm optimization (PSO) to maximize coverage and extend the network lifetime were also proposed in several works. For instance, in [46,47], the VD was used to detect coverage holes and to assess fitness function, while the PSO was utilized to determine the next position of the sensor nodes in a given region of interest. In these studies, a centralized node was employed to collect data about the positions of all nodes, knowing that each node had prior knowledge about its own location as well as those of all other nodes. Then, calculations were done to determine the new locations of each node in the network. The fitness function can be generally applied either to minimize coverage holes or to reduce power consumption.

The literature review demonstrates that few studies combined GA with VD to solve the problem of object collection network deployment and re-allocation. This combination was also successfully used to improve the network coverage and extend its lifetime. However, despite their importance, these approaches applied the fitness function developed by a GA to assess candidate solutions and, then, select the next node positions. Moreover, theVD was only utilized to detect the coverage holes. Another limitation of these methods is that the solutions generated by GA are initialized inside the cells of D V. Despite the importance of these approaches, they have some limitations. In fact, the solutions generated by GA were initialized, in these works, inside the cells of VD. Besides, the nodes constituting the object collection network were put far from or towards theirneighbors to minimize the coverage of the holes [47,42,43]. Moreover, new nodes were sometimes added at the specified locations to cover the holes [48]. Overall, these approaches were performed using knowledge about the locations of the nodes.

## 2.2 Parallel and distributed architecture of GA

As shown in the literature, in GA genetic algorithms, many operators can be executed independently of each other. The efficiency of parallel GAs is to find the desired solution in the shortest time with better performance. Indeed, these algorithms are frequently used to solve large problems and ensure substantial performance gains [49,50]. Most of these algorithms have been run on huge number of parallel machines and their efficiency depends on the parallel computing system. In most of these problems, the ratings of the fitness function can be calculated independently for each candidate solution. In other words, each candidate solution can be computed at the same time (in parallel).

The major objective of using this parallel architecture consists in reducing the execution time and minimizing the number of resources required by GAs. Genetic operations on individuals, such as crossbreeding as well as population mutation and evaluation, can be performed in parallel. The main idea of most parallel programs is to divide tasks into smaller sub-tasks and execute them simultaneously on different nodes. This approach can be applied to genetic algorithms in different ways.

The fundamental difference between these different implementations is to use a single population or to divide it into sub-populations. According to [51], the PGAs (Parallel Genetic Algorithms) can be classified into three main categories as shown in Fig.1:

-The Master-Slave model [52,53]: In this model, there is a master node that manages all sub-populations and distributes the individuals among the slave nodes. Then, the fitness values of the individuals are calculated in the corresponding slave nodes.

-The Coarse-Grained model: It is a distributed model [53] where the population is first divided into many sub-populations located in several "islands". Then, the GA operates independently in each "island". Since each island contains only partial individuals of the population, these islands exchange periodically information by migrating certain individuals to give diversity to the population [54,55]. The model can execute all GA operators in parallel and distributed computing, so that different "islands" can explore periodically various parts of the search space.

-The Fine-Grained model: It is also called "cellular model" or "grid model". It is applied to structure a population into quarters and place some individuals in a node. In this model, the GA is performed in parallel calculation to evaluate the fitness value of each chromosome and apply locally the GA operators. That is to say, selection, crossbreeding and mutation are executed on adjacent neighbors. Based on the massively-parallel architecture, fine-grain model can significantly speed up the evaluation of all chromosomes. However, this cannot be done only by using a massive clustering system to manage this model [56].

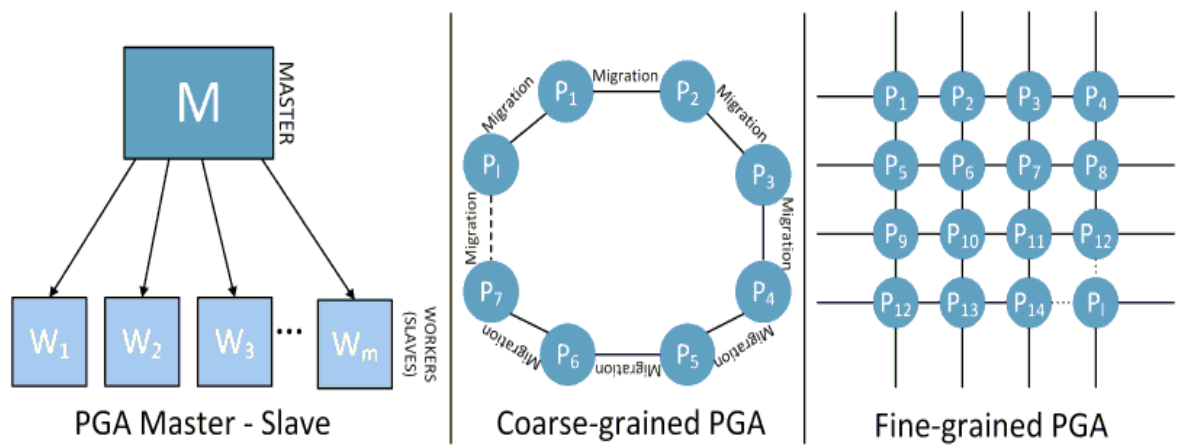

Fig.1 Different types of parallel Genetic Algorithms [57]

The objective of this study is to design a distributed approach where the hybridization of the Voronoi diagram and the GA is executed in parallel in several nodes. The VD is essentially used to divide the region of interest into cells and generate initial solutions that present the initial population of the GA. Then, this population will be divided into in sub-populations which will be, later, distributed to the different nodes constituting the employed data collection network. Afterwards, a genetic algorithm will be executed in parallel in each of these nodes. Each node will return its best solution having the maximum coverage in a RoI.

## 3 Research Design

### 3.1 Distributed strategy

In our proposed approach (Fig. 2), two types of nodes are used: Voronoi nodes and Genetic nodes. The former (V.Node) generate solutions presenting the initial population which will be subdivided into sub-populations and send them to genetic nodes (G.Node). The hybridization of the two algorithms is done mainly at each G.Node. Subsequently, the fitness values of individuals are calculated in each G.Node to improve coverage. Finally, the found optimal solutions will be returned to the V.Node. As the suggested approach is based on two-way communication, it cannot be the Master-Slave model.

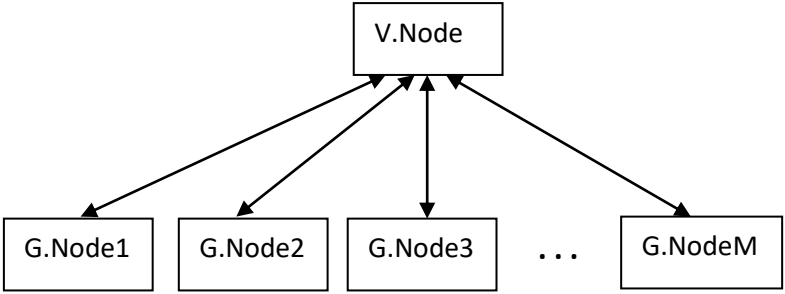

Fig.2 Proposed distributed architecture

Our major aim consists in developing a protocol that allows managing and carrying out communication between the different nodes in the approach presented in Fig. 2. This protocol is the core of the "ESP-NOW" protocol that presents a library under Arduino. According to [58], ESP-NOW, permits several electronic cards to communicate with each other based on direct links between nodes to form an ad-hoc network without using a Wi-Fi access point as an infrastructure. In fact, it is similar to high-speed wireless connectivity which allows the exchange of small 2.4GHz frequency bands data packets with low consumption in ESP-NOW.

To develop our protocol and design our distributed approach, we used the same type of nodes "M5StickC ESP32-PICO Mini IoT Development Kit" [59].

The introduced approach is based on dividing the initial population into smaller sub-populations and simultaneously performing the corresponding tasks on different nodes. For example, a population of 300 solutions generated by V.Node is distributed over 10 nodes (G.Nodes); each of which implements a population of 30 individuals. According to Fig. 3,the proposed protocol is utilized in four types of frames; each frame has a payload of up to 250 bytes that can be transported:

- SPF: A Frame-Sub-Population contains M Voronoi solutions and it is sent from V.Node to all G.Nodes
- RF: Result-Frame includes the optimal solution of each node. After the execution of its algorithm, each G.Node returns a RF frame to the V.Node.
- FPF: Final-Position-Frame involves the final position of each node. The V.Node gathers all the RF frames returned by each G.Node and, subsequently, selects the best solution which presents the maximum obtained coverage.
- ACK: or Acknowledgment, transmitting in both directions (Fig.3), used to inform the application layer about the transmission success or failure of each type of frame.

The greatest amount of information is observed at the SPF frame. In this case, segmentation can be carried out at the TCP / IP layer; hence the SPF frame will be subdivided into sub-frames.

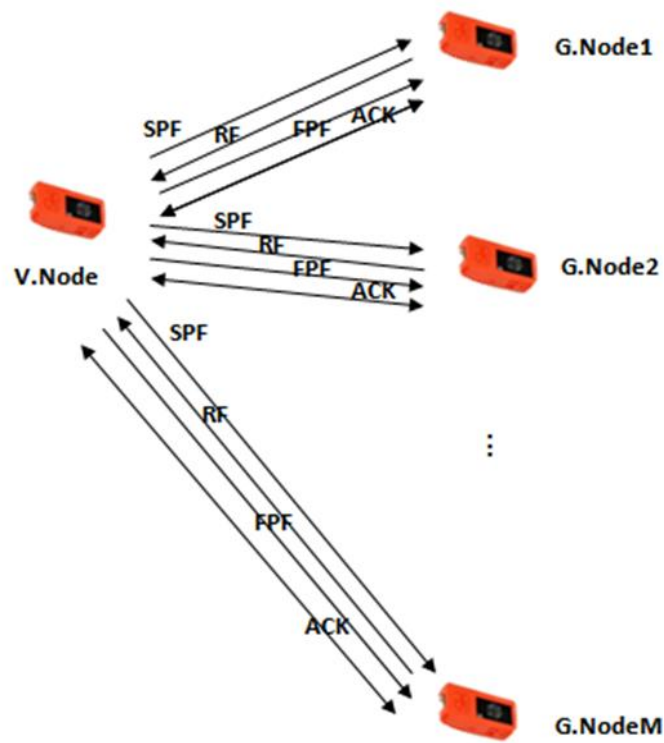

Fig.3 Proposed network architecture

Another criterion of the proposed protocol consists in automatically connecting one of the used devices to its counterpart to continue communication if it is suddenly reset or if there is a loss of power when it restarts.

Fig. 4 shows a sequence diagram of our communication protocol applied on three nodes; each of which has a transmission role and a reception role (T / R) data frame. The V.Node in the middle and the two G. Nodes are on the left and right. At the instant t = 0, the V. Node generates first the

Voronoi solutions and, then, distributes them to the different G.Nodes. In this example, the generated Voronoi solutions are distributed over 6 SPF frames. The first three frames are those of G.Node n° 1, while the remaining frames are those of G.Node n ° 2. The sending delay of each frame is equal to 10ms. In fact, each G. Node receives its SPF frames that present the initial population to execute its VD-GA algorithm. After execution, each G.Node sends the found solution in an RF frame. The V.Node always "listening", gathers all the RF and applies the best found solution (the solution having the best coverage). This solution will be disseminated on G.Node by the FPF frame.

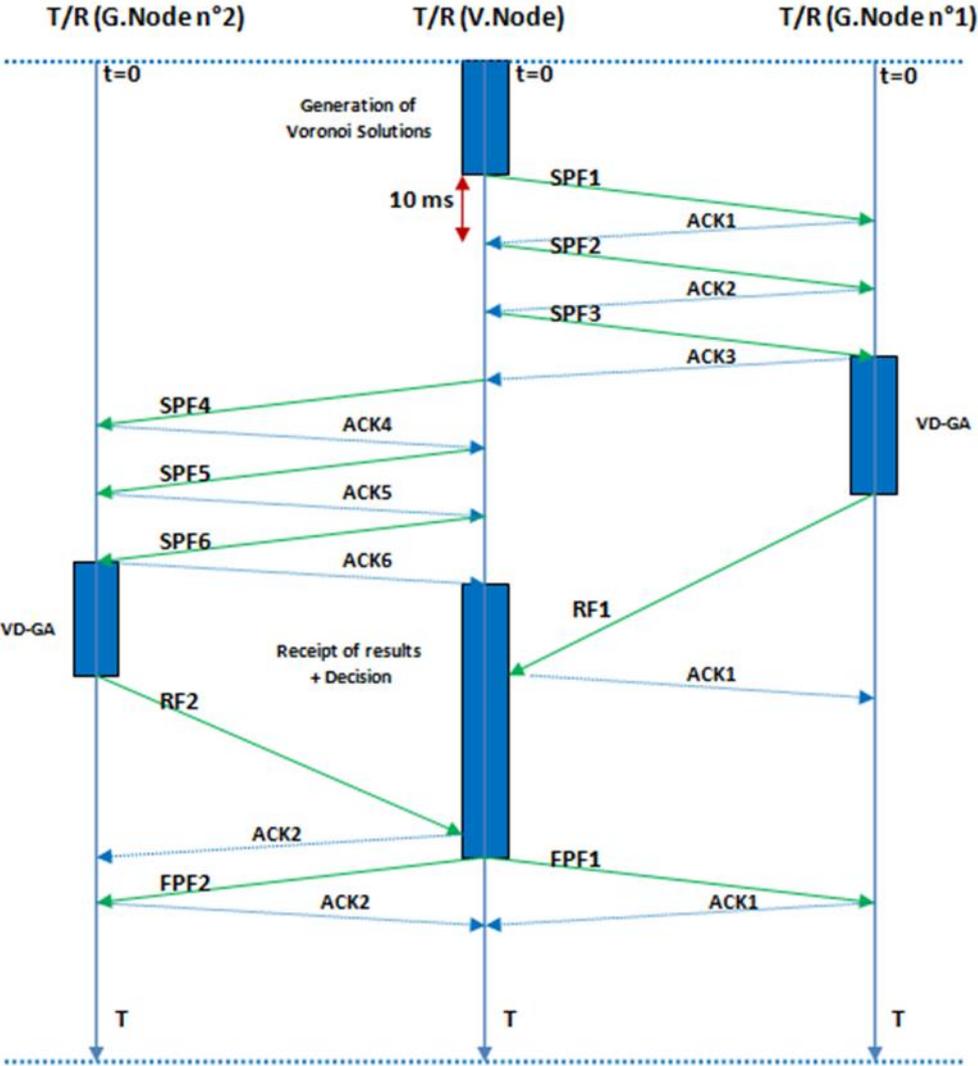

Fig.4 Sequence diagram of the proposed protocol of communication between three nodes

## 3.2 Design of the distributed positioning algorithm

### 3.2.1 Deployment Scenario

As shown in Fig.5 the introduced approach starts by generating the Voronoi solutions in V.Node that subdivides the latter into sub-populations; each of which will be assigned to a G.Node. The process independently develops its populationuntil it decides to bring together its best individuals who will be candidates at the decision stage.

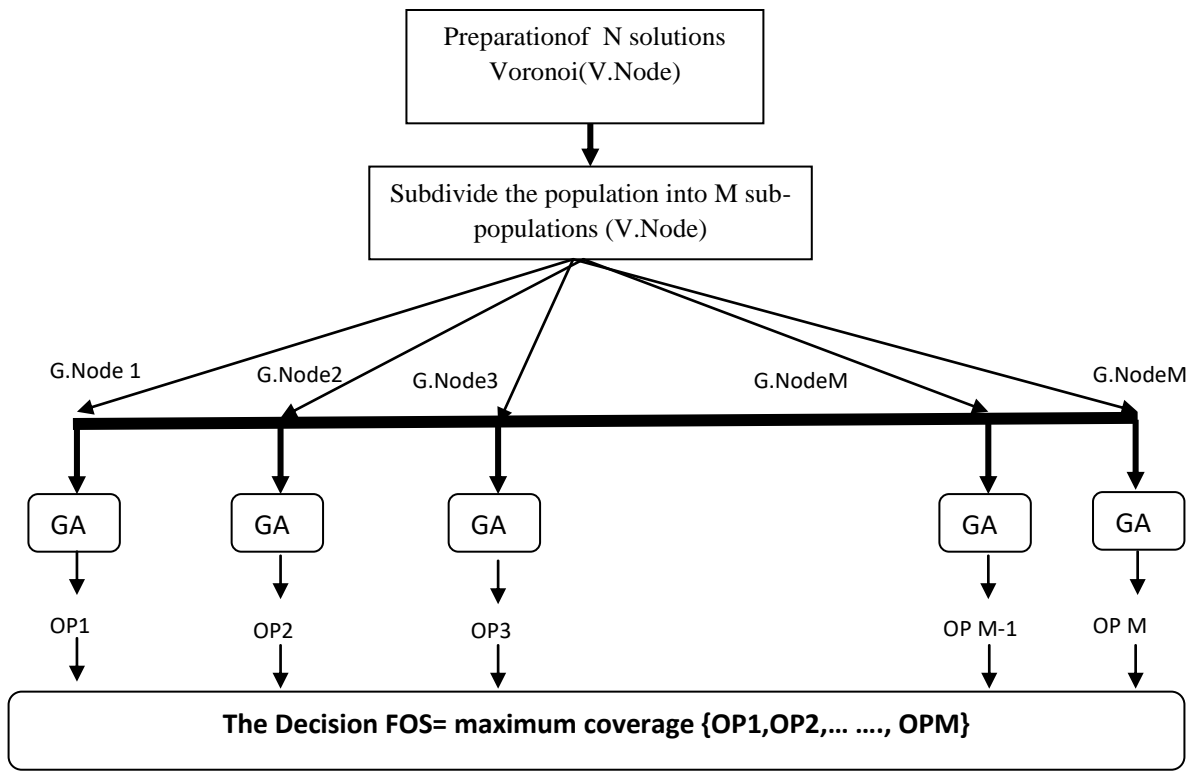

Fig.5.Functioning of the proposed approach

### 3.2.2 Deployment scenario: A hybrid optimization algorithm

This algorithm starts with implementing a Voronoi diagram to generate initial solutions (Pop initial) and, then, subdividing it into sub-populations (lines 2 and 3). Afterwards, the genetic algorithm is executed in parallel on all G.Nodes until a desired maximum coverage will be obtained (from line 3 to line 7 of the algorithm). The inputs of the algorithm are: M (the number of nodes to deploy), N-solutions (the number of solutions to generate by V.Node), dimension of region of interest, pop_size (the size of the population), the desired coverage (stop condition) or a given number of iterations.

**Algorithm 1: VD-GA Voronoi-Genetic**.

**0) Define the evaluation function f (Maximum desired coverage);**

**1) Pop $_{initial}$ ← N solutions generated by the Voronoi diagram;**

**2) Subdivide the initial population initial (Pop)**

**3) Do it in parallel**

**4) Apply the AG on each G.Node Node (M Voronoi solutions for each node):**

**Repeat**

a) Selection
b) Crossbreeding
c) Mutation

UNTIL the coverage will be converged

5) Find the best solution among all solutions generated by each node

6) Go to 3 or Stop,

7) End of parallelism,

8) Return the best Solution.

9) Assignment of the final position

**Voronoi Diagram**
Generally, VDs are used in networking to find collision-free paths. Therefore, they are employed to construct paths in cartes. To deploy node in data collection networks, we utilized the VD to generate initial solutions presenting the positions of the deployed IoT objects.
Let S be a finite set of n points on the map (as shown in Figure 6), a Voronoi region associated with an element p of S is the set of points of S which are closest to p.
$$Vors(p) = \{x \in \mathbb{R}^2 \; \forall \, q \in S \|x - p\| \leq \|x - q\|\} \quad (1)$$

Where $\|x - p\|$ is the distance between $x$ and $p$.

**Genetic algorithm**
In this paper, we apply the standard GA with some adequate modifications explained in the following sections. The GA process begins with a set of individuals constituting a set of randomly-generated possible solutions called population. This genetic process is repeated until a stop condition is met.

- **Chromosomes** coding: This step aims at presenting an individual that presents a solution (the position of each node in a given region of interest). Fig. 6 illustratesa chromosome showing a feasible solution in the search space, an example of individual representation, with n = 6 objects and k = 1 (a single type of node).

| Object1 | Object2 | Object 3 | Object 4 | Object5 | Object 6 |
|---------|---------|----------|----------|---------|----------|
| $(x_1,y_1)$ | $(x_2,y_2)$ | $(x_3,y_3)$ | $(x_4,y_4)$ | $(x_5,y_5)$ | $(x_6,y_6)$ |

Fig.6. Representation of a chromosome in the GA

- **Initial Population**: The GA process generally begins with an initial population which presents the set of all individuals used to find the global solution. In our approach, the initial population presents a sub population received and generated by the V.Node (the node which will generate the initial solutions).

- **Fitness Function**: The objective of our approach is to determine the best positions that guarantee the maximum coverage of RoI. The fitness function is usually applied in GA to assess and identify the best-found solutions.

$$\text{Fitness} = S - \sum_{i=0}^{n} surfaces of all circles \quad (1)$$

Where S denotes the region of interest (RoI) area and n is the number of deployed objects. The more the difference between S and the sum of all the areas of the objects decreases, the more the coverage increases. If there is no overlap between two objects, the distance separating them will be equal to or greater than 2r, as exposed in Figure 7 (b):

$$X = 2r - distance\ (p1, p2) \quad (2)$$

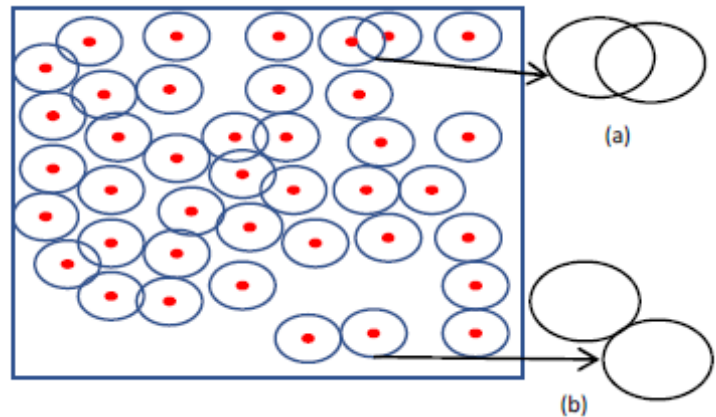

Fig.7. Objects overlap [37]

- **Selection of** individuals: It consists in selecting two chromosomes as parents of the population to produce two new individuals.
- **Crossbreeding**: A random point in the two selected individuals is chosen to exchange genes after this point.
- **Mutation**: It consists in randomly choosing and mutating a gene in our approach, i.e. changing the position of a node within the individual. The mutation guarantees diversity and avoids premature convergence by sufficiently exploring the research space to bring innovation to the population.
- **Stopping condition**: Two stopping criteria were used in our experiments: the coverage rate and the maximum number of generations.

## 4 Results and discussions

To evaluate the introduced distributed approach of node deployment by applying our VD-GA algorithm in a field of interest, a real prototyping is suggested. It was tested in a real environment with experiments carried out on a testbed containing of 6 to 10 nodes. Before this prototyping, we started with simulations.

### 4.1 Simulations

To evaluate our VD-GA algorithm with a distributed approach, we used OMNET ++ in the simulations. First, the DV was applied on a RoI of 80 * 80 meter. The DV divides this region into a set of Voronoi cells to generate a population of 100 solutions (individuals) that present the positions of each IoT object in a 2D plan. Then, this population was divided into sub-populations over 20 nodes; each of which has a detection range of 10 meters. The choice of the parameter values is based on empirical tests and on our previous studies presented in [37]. Table 2 shows the used simulation parameters.

Table2. The simulations parameters

| Parameters | |
| --- | --- |
| Population size: Number of individuals | 100 |
| Number of iterations | [100..1500] |
| Number of deployed objects | 20 |
| Crossover rate | 0.9 |
| Mutation rate | 0.1 |
| Size ROI | 80*80 |
| Stopping Condition | 94% |

Subsequently, the GA was executed on each node in parallel to optimize the solutions found by the VD by replacing the nodes in order to find the best positioning for them in the RoI and, subsequently, to obtain the maximum degree of coverage. Figures 8 and 9 show the difference, in terms of the degree of coverage, between the initial random coverage, generated only by the Voronoi diagram, and the degree of coverage provided by the proposed VD-GA.

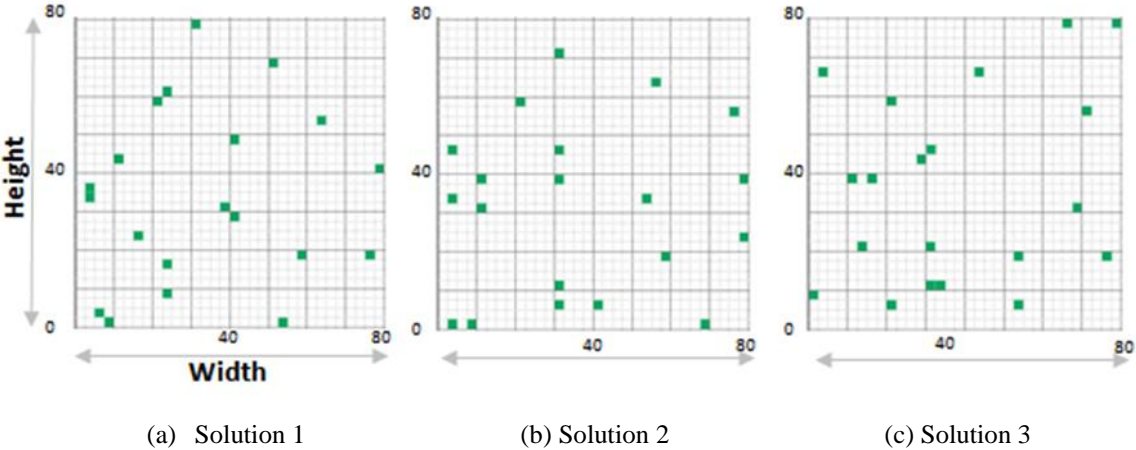

(a) Solution 1　　　　　　(b) Solution 2　　　　　　(c) Solution 3

Fig.8. Deployment of 20 nodes by VD (example of 3 solutions)

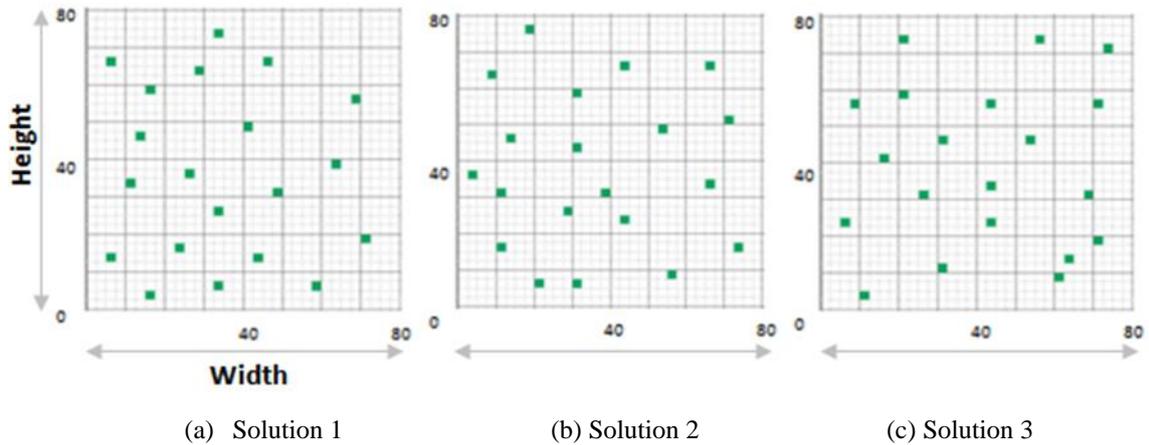

(a) Solution 1          (b) Solution 2          (c) Solution 3

Fig.9. Deployment of 20 nodes after executing VD-GA (example of 3 solutions)

Fig. 10 clearly reveals the difference in terms of coverage between random deployment, deployment using only the VD, deployment utilizing only the GA and deployment of hybridization between VD and GA. It is obvious, in this figure, that the degree of coverage is improved as the number of GA iterations increases. It is also noticed that the GA outperforms the DV for a number of GA iterations inferior to 100. On the other hand, the GA gives a better degree of coverage during more than 100 iterations. The hybrid (VD-GA) is more efficient than VD and GA, if used separately during a number of GA iterations exceeding 56.

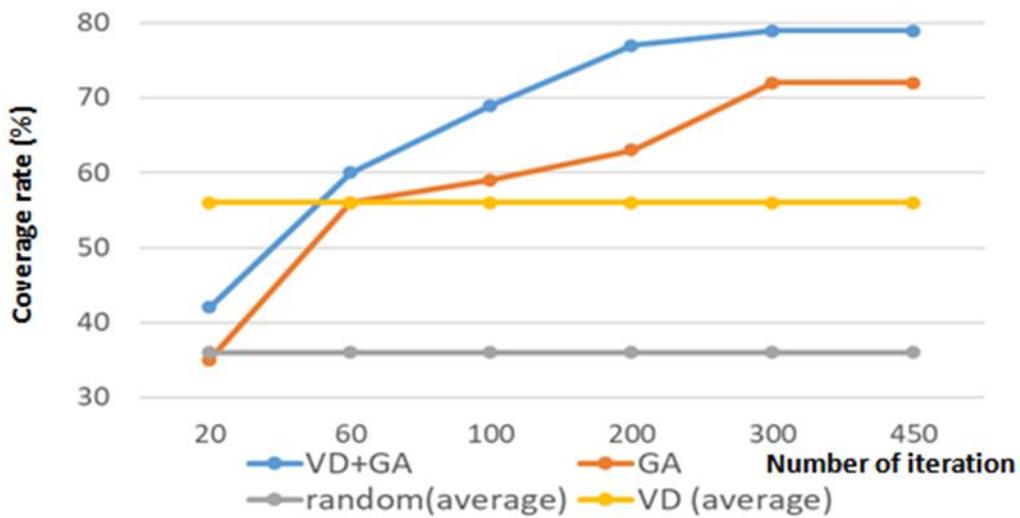

Fig.10. Comparison of the coverage degrees according to the Number of iterations for VD, GA and VD-GA

Figure11 shows the increase of the coverage degree after each iteration in VD-GA.

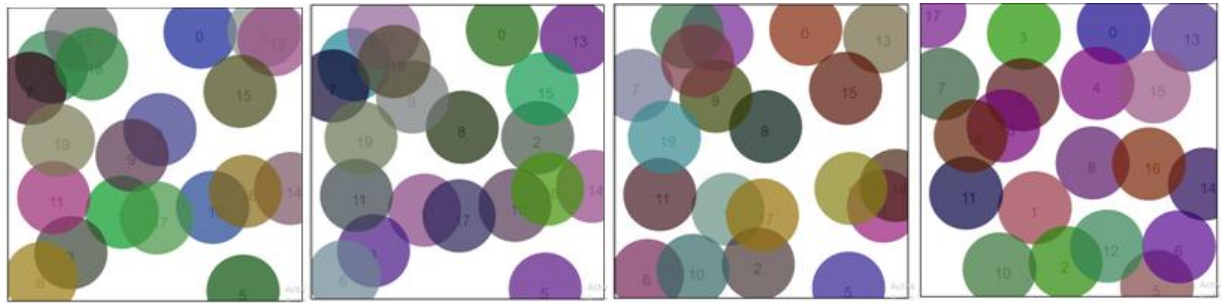

| Iteration 1: 61% | Iteration 20: 72% | Iteration 100: 83% | Iteration 197: 91% |

Fig.11 Distribution of 20 nodes in a 80*80 region

Fig. 12 shows the evolution of the GA execution over the first 15 iterations. This algorithm started with 58% coverage. After the operations of selection, crossbreeding (between Parent 1 and Parent 2) and mutation, the coverage could be calculated by applying the fitness function. In fact, if this degree of coverage increased, the new parents would be Fetus 1 and Fetus 2. Otherwise, a new selection will be executed.

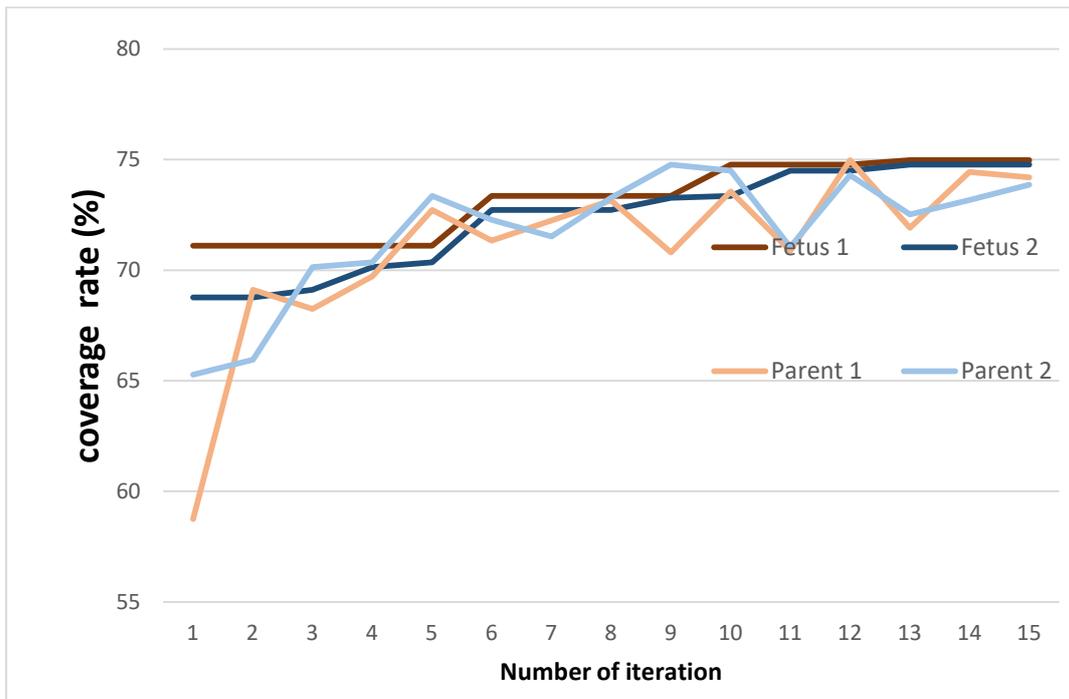

Fig.12. Functioning of GA in each node

### 4.2 Testbeds setting

A real testbed implementing several real nodes was designed and organized in a wireless network. Only one type of node was used in our experiments; it is the M5StickC node of the M5Stack family powered by ESP32, are electronic cards equipped with a 4 MB Flash memory, a 2.4G antenna, an IR transmitter, a microphone, buttons, LCD screen (0.96 inch), and built-in Lipo Battery as demonstrated in Fig. 13. M5StickC nodes can be developed on UIFlow, MicroPython and Arduino platforms.

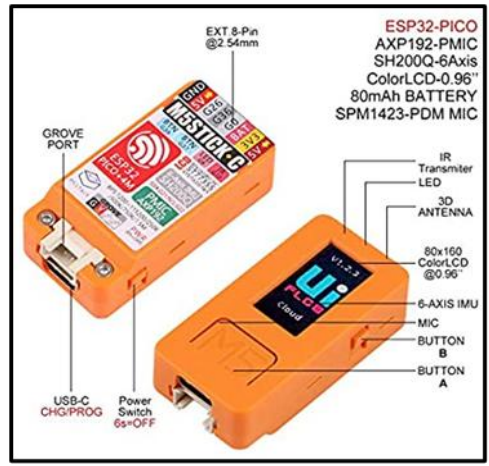

Fig.13. The used M5StickC ESP32 Mini IoT nodes [59]

Tab.3 presents the used experimentation parameters

Table3. The experimental parameters

| Parameters | |
|---|---|
| Population size: Number of individuals | 100 |
| Number of iterations | [400..1500] |
| Number of deployed objects | 06 |
| Crossover rate | 0.9 |
| Mutation rate | 0.1 |
| Size of RoI | 25* 17 (2 floor) |
| Stopping conditions | 94% |
| Threshold of the received result | 80% |

06 M5StickC nodes were deployed in an apartment composed of a ground floor and 2 other floors; each has a surface area of 25 * 17 meters. The ground floor and the first floor contain separately six bedrooms, a wall separation whose thickness ranges from 20 to 30 cm between 2 neighboringbedrooms (as shown in Fig.14) and a 30-cm concrete ceiling between the two floors. The second floor is a free space. The transmission range of the nodes was in an interval of 10 to 18 meters, depending on the obstacles encountered during transmission. The RSSI (Received Signal Strength Indicator) and FER (Frame Error Rates) were calculated. The transmission power of Wi-Fi was 100 mW. Due to the stochastic nature of optimization algorithms, the optimization process was performed 30 times and an average value was computed for all values presented in the next figures.

We noticed that most G.Nodes found their optimal solutions within the same timeframe, while there are other nodes which take longer time for data recovery and decision making to seek the best solution. Since our stop condition was defined by a desired coverage, a new parameter

"Received result threshold" was added. For example, if 10 nodes were deployed and if the result threshold was set to 80%, the V.Node would force the process to stop and take the decision as soon as it received the results of 8 G.Node.

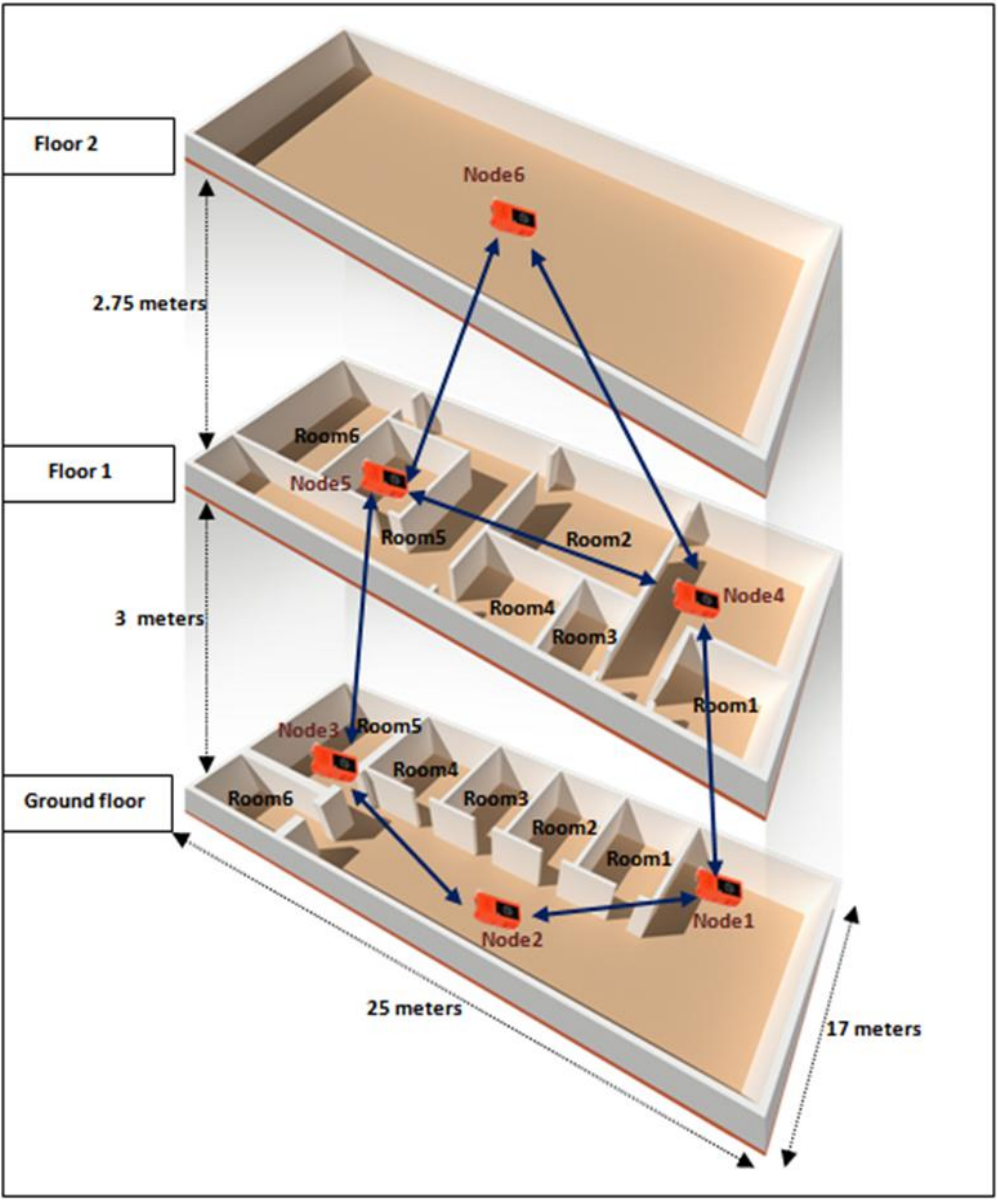

Fig.14. RoI of experimental tests

**4.3 Results of evaluation of the proposed GA-VD algorithm**

In this section, we compare the results obtained by the distributed approach proposed in the present work with those provided by a centralized approachin terms of percentage of found coverage and execution time. Moreover, the execution of GA, where we can use the mutation operation twice in each generation, is compared to use one mutation Third comparison on energy consumption in the two approaches. Fourth comparison with the average RSSI rates and the number of neighbors of the nodes. These results were provided with an execution average of 40 times.

**4.3.1 Comparisons of the centralized approach and our distributed approach**

We re-evaluated our previous study [37] on a single M5StickC node. That is to say, the allprocessingwas performed on a single node; hence the use of a centralized approach based on a single population generated by the DV and combined with GA.

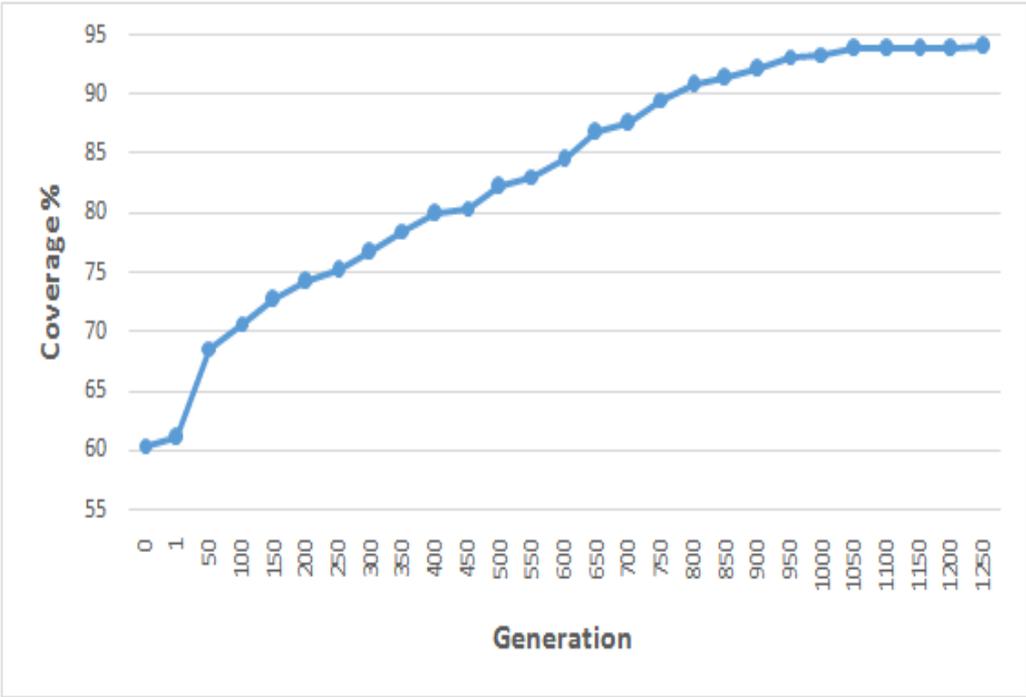

Fig.15. Coverage rate obtained by the centralized approach

As revealed in Fig.15, 94% coverage was provided by the centralized approach in the $1250^{th}$ iteration.

Fig. 16 shows the coverage rate of six nodes used in a distributed approach. Obviously, the nodes achieved their desired coverage (94%) at different numbers of iterations. For instance, node 4 attained its coverage in the $706^{th}$ iteration, while node 5 found its cover in iteration number 952.

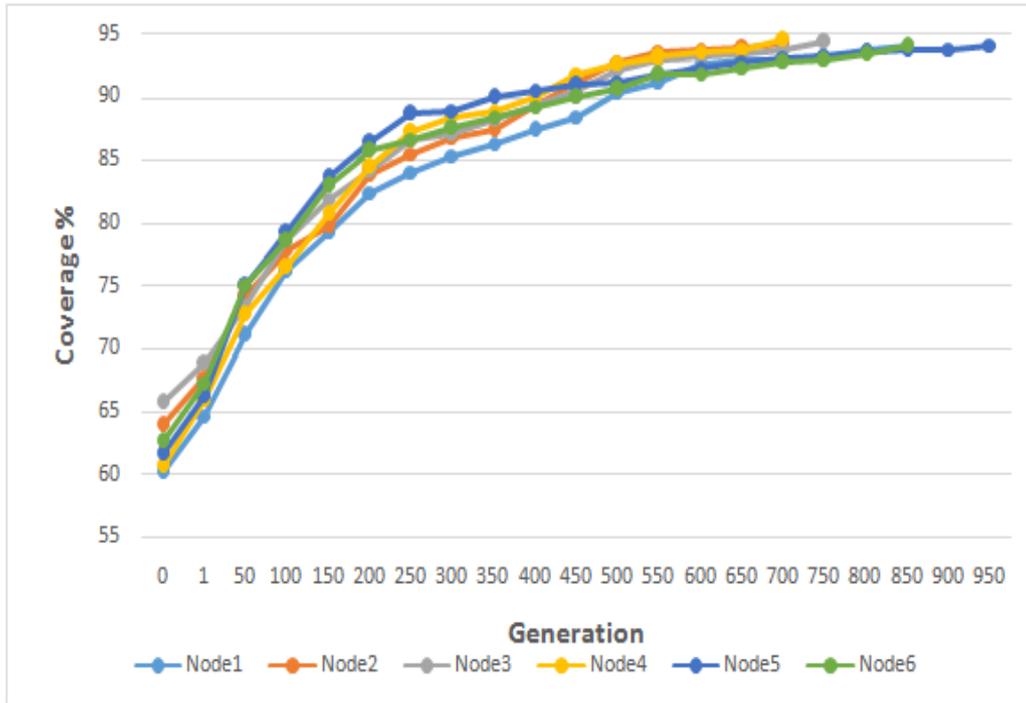

Fig.16. Coverage rate obtained by the distributed approach for the six nodes

The objective of these evaluations is to compare the coverage rate of the distributed approach with that of the centralized one.

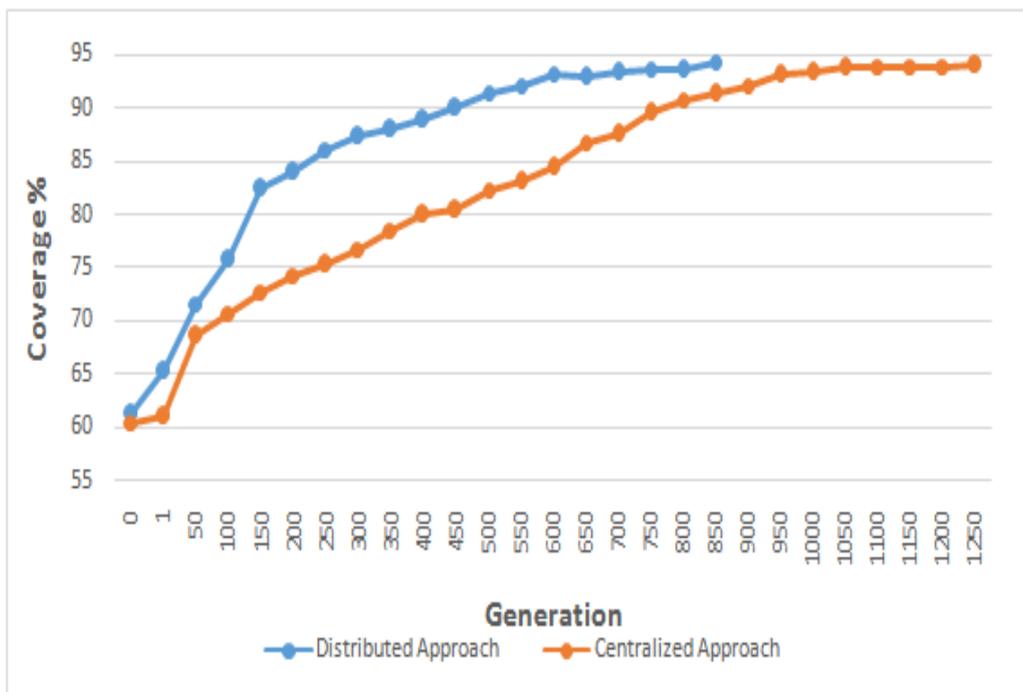

Fig.17. Comparison of the coverage rate obtained by the distributed approach with that provided by the centralized approach

We measured the percentage of surface coverage in the 2 approaches with a stop condition of 94%. In the distributed approach, we summed the averages of the results received from the different nodes that reached this coverage rate. We can clearly notice that a distributed approach reached the goal in the 800thgeneration, while the centralized approach continued to run for 450 moreiterations,as illustrated in Fig.17.

Although in a distributed approach, there is a computational time cost (e.g. the time required to distribute the sub-populations to each node, the execution time and the time of returning the found solution), the graph presented in Fig.18 clearly shows that the distributed approach ended its operation in shorter time than that required by a centralized approach. This result can be explained by the cost of calculating the fitness function ofeach individual in a large population used in the centralized approach.

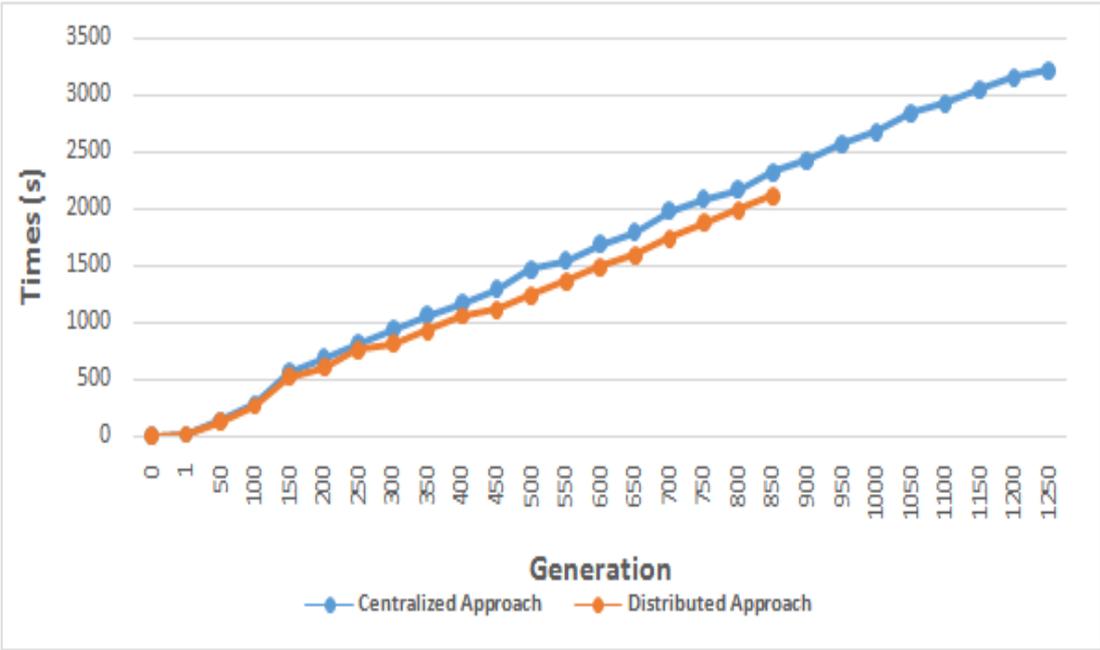

Fig.18. Average computing time required by the centralized approach and distributed one

### 4.3.2 Evaluation of the proposed algorithm in terms of energy consumption

Energy consumption is a parameter widely used to compare the two approaches. For this reason, in the conducted experiments, all nodes were fully charged (100%) at the same time. In the distributed approach, we calculated the average energy consumption of all nodes used in this approach. Tab.4 presents a comparison of the two approaches in terms of level of charge of battery. It also shows that an M5StickC node in the centralized approach consumed approximately 54% of the total energy until it stopped operating, versus 28% for the distributed approach.

Table4. Battery charge level

| Times (S) | 0 | 500 | 1000 | 1500 | 2000 | 2500 | 3000 |
|---|---|---|---|---|---|---|---|
| Centralized Approach | 100% | 96% | 86% | 71% | 65% | 57% | 46% |
| Distributed Approach | 100% | 96% | 83% | 77% | 72% | ------- | ------ |

**4.3.3 Evaluation of the proposed algorithm with double mutation**

In a GA, after the selection and crossbreeding operations, each individual will participate in a mutation at a given time point. Usually, the location of the mutant gene that will be replaced with a different value is random. In this section, we compare the result provided by the proposed VD-GA hybridization in a distributed approach, which uses a single mutation, with those obtained by double mutation, i.e. instead of changing the position of a single node, we change the position of both nodes at the same time.

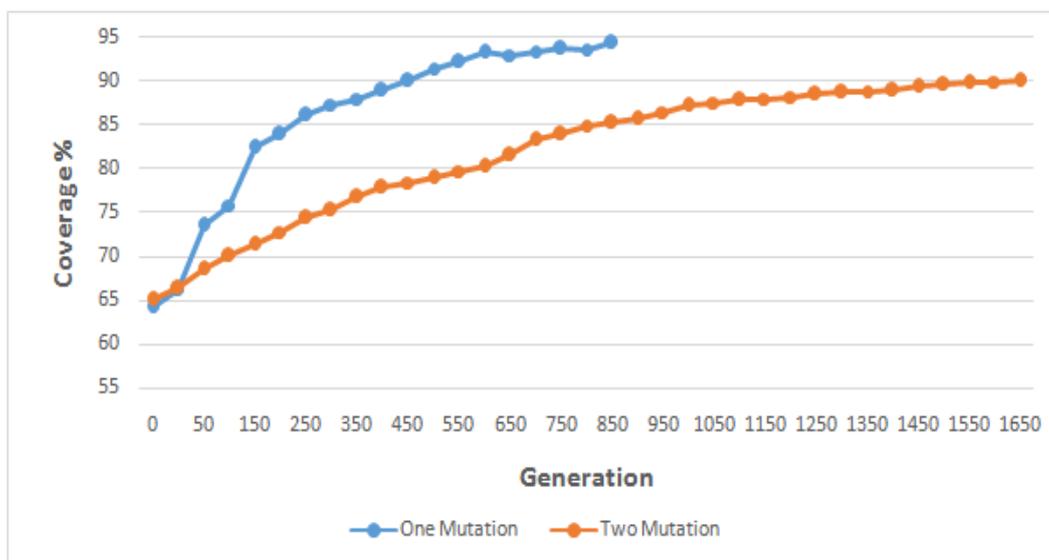

Fig.19. One-point and two-point mutation operator

Figure 19 illustrates an example of deploying 10 nodes in a 2D plane. For example, after the executing a single mutation, node 5, which is at position (84,29), will be at position (24,53). For the double mutation, two objects will be repositioned, node 2 (66, 73) will be at position (29,79) and, at the same time, node 8 (18,47) will be at position (68,32).

After applying the introduced VD-GA hybridization in a distributed approach, a GA was used with a single mutation ending after about 800 iterations. However, a GA with two mutations required more than 750 iterations to achieve 90% of surface coverage, as revealed in Fig. 20.

Fig.20. Comparison of one-point and two-point mutation operator

**4.3.4 Evaluation of the genetic-Voronoi algorithm according to network metrics (RSSI, lifetime and number of neighbors)**

Tables 5, 6 and 7 represent the average values of RSSI, lifetime and number of neighbors of the used nodes.

Table5. Average values of number neighbors for 6 nodes

| Node | VD | GA | VD+GA |
|---|---|---|---|
| 1 | 132.39 | 131.83 | 137.28 |
| 2 | 135.28 | 128.92 | 136.64 |
| 3 | 131.23 | 125.17 | 131.55 |
| 4 | 114.72 | 116.28 | 115.21 |
| 5 | 89.29 | 92.88 | 91.53 |
| 6 | 129.82 | 127.76 | 129.16 |

RSSI or Received Signal Strength Indicator is a measure of a device's ability to receive a signal from an access point or router. This value helps determine if there is enough signal to establish a good wireless connection. The RSSI value is retrieved from the client device's Wi-Fi card, it is not the same as the transmit power from the router or AP.

Table 5 shows high values of RSSI issued from the nodes positioned according to the proposed hybrid genetic-Voronoi algorithm (VD+GA). The superiority of these RSSI values indicated a better ability of localization of nodes using the genetic-Voronoi algorithm.

Table6. Average values of lifetime (in seconds) for 6 nodes

| Node | VD | GA | VD+GA |
|---|---|---|---|
| 1 | 1298 | 1367 | 1693 |
| 2 | 987 | 927 | 996 |
| 3 | 1029 | 1035 | 1083 |
| 4 | 703 | 749 | 753 |
| 5 | 870 | 883 | 892 |
| 6 | 943 | 932 | 921 |

Table 6 illustrates high lifetimes of nodes positioned according to the proposed hybrid genetic-Voronoi algorithm (VD+GA). This indicated a better ability of coverage (deployment) of nodes using the genetic-Voronoi algorithm.

Table7. Average values of number neighbors for 6 nodes

| Node | VD | GA | VD+GA |
|---|---|---|---|
| 1 | 4.23 | 4.65 | 4.98 |
| 2 | 3.22 | 3.41 | 3.56 |
| 3 | 2.68 | 2.91 | 3.02 |
| 4 | 3.56 | 3.86 | 3.94 |
| 5 | 4.29 | 4.44 | 4.62 |
| 6 | 3.12 | 3.42 | 3.68 |

Table 7 indicates high numbers of neighbors of the nodes deployed according to the suggested VD+GA. The superiority of this number of neighbors reflects the better quality of connectivity of nodes using the genetic-Voronoi algorithm.

It can be deduced from tables 5, 6 and 7 that the proposed VG+GA algorithm achieved better results regarding highest RSSI average values, highest lifetime average values and highest number of neighbors average values. This can be explained by the advantage of the hybridization and the distribution. Indeed, the hybrid model takes advantages of the both genetic and Voronoi process in finding the most appropriate solutions for the repartition of the nodes. Moreover, the distribution allows a better managements of the computation tasks when calculating the costs of possible solutions and its corresponding evaluation network metrics (RSSI, ..) which, in turn, ameliorate the quality of solutions.

### 4.4 Discussion and interpretations

Our VD-GA hybridization with a distributed approach shows better coverage, compared to a the centralized approach. Our proposal used fewer nodes with better coverage, compared to other results [41,42]. Moreover, a distributed approach with a GA that uses a single mutation is more efficient in terms of time calculation cost than a GA that uses two mutations.

By comparing the simulations, presented Fig.12 Fig.13 and Fig14, and the real experiments, shown in Fig.18, we notice that the distances separating some nodes are not the same. This difference can be explained by the existence of obstacles between two neighboring nodes in a real prototyping environment. The localization error obtained in the conducted indoor experiments was of the order of 20 cm, which is generally acceptable for indoor control systems and does not affect the results.

### 5 Conclusions and perspectives

In this stydy, the problem of deploying objects in an indoor data collection network was studied. The proposed approach was used in two contributions: The first was to combine between a voronoi diagram and the genetic algorithm, while the second is to design a distributed approach.VD was also applied to randomly generate the locations of IoT objects.Theselocations present the initial population that would be subdivided into sub-populations; each of which was assigned to a G.Node. The GA was applied to determine the best locations of objects. Compared to other approaches, that proposed in this manuscript allowed enhancingthe deployment solution. The only encountered obstacles is that the memory size of an M5StickC node did not allow our approach to be executed when the number of nodes to be deployed exceeded 58 nodes. As a future study, the aim is to propose other models of hybridizations of the geometric deployment methods (other than VD) and optimization algorithms (other than GA).In addition, another future research direction is to evaluate the behavior of the proposed VD-GA on large-scale experimental environments in which the number of nodes exceeds 100.Since hybrid models generally have high complexities [60], we propose to establish an analysis of the temporal and algorithmic complexity and a statistical study in order to prove the scalability of the proposed model.